\documentclass{article} %
\usepackage[final]{neurips}

\usepackage{xcolor}         %
\usepackage{xspace}         %
\usepackage{colortbl}
\usepackage{color}
\usepackage{microtype}
\usepackage{hyperref}
\usepackage{url}
\usepackage{tcolorbox}
\usepackage{booktabs}
\usepackage{times}
\usepackage{latexsym}
\usepackage{graphicx}
\usepackage{booktabs}
\usepackage{fancyvrb}
\usepackage{wrapfig}
\usepackage{enumerate}
\usepackage[T1]{fontenc}
\usepackage{highlight}

\usepackage{soul}
\usepackage{tipa}

\usepackage[utf8]{inputenc} %
\usepackage[T1]{fontenc}    %
\usepackage{hyperref}       %
\usepackage{url}            %
\usepackage{booktabs}       %
\usepackage{amsfonts}       %
\usepackage{nicefrac}       %
\usepackage{microtype}      %
\usepackage{xcolor}         %

\usepackage{amsmath}
\usepackage{multirow}
\usepackage{array}
\usepackage{siunitx}
\usepackage{booktabs}
\usepackage{comment}
\usepackage{tikz}
\usepackage{listings}
\definecolor{pos}{RGB}{167, 199, 231}
\definecolor{neg}{RGB}{250, 160, 160}

\usepackage{inconsolata}

\newcommand{\metricrank}{$p$-MRR}

\definecolor{darkblue}{rgb}{0, 0, 0.5}
\hypersetup{colorlinks=true, citecolor=darkblue, linkcolor=darkblue, urlcolor=darkblue}

\author{
    \textbf{Orion Weller}$^{\hspace{.1em}
    \hspace{.1em}{\color{blue}\boldsymbol{\iota}}}$
    \quad
    \textbf{Benjamin Chang}$^{\hspace{.1em}\color{blue}\boldsymbol{\iota}}$
    \quad
    \textbf{Sean MacAvaney}$^{\hspace{.1em}\color{blue}\boldsymbol{\lambda}}$
    \quad
    \vspace{.2em}
    \textbf{Kyle Lo}$^{\hspace{.1em}\color{blue}\boldsymbol{\alpha}}$
    \quad \\
    \textbf{Arman Cohan}$^{\hspace{.1em}\color{blue}\boldsymbol{\gamma\hspace{.1em}\alpha}}$
    \quad
    \textbf{Benjamin Van Durme}$^{\hspace{.1em}\color{blue}\boldsymbol{\iota}}$
    \quad
    \textbf{Dawn Lawrie}$^{\hspace{.1em}\color{blue}\boldsymbol{\iota}}$
    \quad
    \textbf{Luca Soldaini}$^{\hspace{.1em}\color{blue}\boldsymbol{\alpha}}$
    \vspace{.5em}\\
    $^{\color{blue}\iota\hspace{.1em}}$Johns Hopkins University
    \quad
    $^{\color{blue}\alpha\hspace{.1em}}$Allen Institute for AI
     \vspace{.5em}\\
    $^{\color{blue}\lambda\hspace{.1em}}$University of Glasgow
        \quad
    $^{\color{blue}\gamma\hspace{.1em}}$Yale University
    \vspace{.5em}\\
    \texttt{oweller@cs.jhu.edu}
}

\begin{document}

\newcommand{\dataset}{\textsc{FollowIR}}

\title{\dataset: Evaluating and Teaching Information Retrieval Models to Follow Instructions}
\maketitle

\begin{abstract}
Modern Language Models (LMs) are capable of following long and complex instructions that enable 
a large and diverse set of user requests.
While Information Retrieval (IR) models use these LMs as the backbone of their architectures, 
virtually none of them allow users to provide detailed instructions alongside queries, thus limiting their ability to satisfy complex information needs.
In this work, we study the use of instructions in IR systems. 
First, we introduce our dataset \dataset, which contains a rigorous instruction evaluation benchmark as well as a training set for helping IR models learn to better follow real-world instructions.
\dataset\ repurposes detailed instructions---also known as \textit{narratives}---developed for professional assessors to evaluate retrieval systems.
In particular, we build our benchmark from three collections curated for  shared tasks at the Text REtrieval Conference (TREC).
These collections contains hundreds to thousands of labeled documents per query, making them suitable for our exploration.
Through this process, we can measure how well IR models follow instructions, through a new pairwise evaluation framework.
Our results indicate that existing retrieval models fail to correctly use instructions, using them for basic keywords and struggling to understand long-form information.
However, we show that it is possible for IR models to learn to follow complex instructions: our new \dataset-7B model has significant improvements after fine-tuning on our training set.
\end{abstract}

\section{Introduction}
Modern language models (LMs) are extensively tuned to be able to follow user instructions faithfully~\citep{chung2022scaling,instructGPT,dpo,wang2023far,ivison2023camels} and safely~\citep{anthropic-hh,bianchi2024safetytuned}.
Through these capabilities, LMs are able to successfully tackle a broad range of tasks~\citep{chiang2024chatbot,liang2023holistic,gpt4tools,jimenez2024swebench,zeng2023evaluating}, even when not explicitly fine-tuned for them.

In contrast to the broader LM community, information retrieval (IR) practitioners and researchers have yet to fully exploit instruction-tuned models.
Thanks to their ability to effectively estimate semantic similarity between query and documents, LMs have been adopted as the main backbone of neural retrieval architectures~\citep{karpukhin2020dense,khattab2020colbert,reimers2019sentence}.
However, the vast majority of these systems are fine-tuned to operate exclusively as text spans similarity estimators~\citep{khattab2020colbert,izacard2021unsupervised,nogueira2019passage,pradeep2023rankzephyr,ma2023fine}.
Moving past these \textit{ad-hoc} search systems to retrieve \textit{with instructions} would enable support for complex information needs.
For example, imagine a researcher seeking to identify papers that must contain numerous qualities to be relevant (from a given venue, using a particular class of methods, etc.) while also making sure to avoid conditions that would make the document not-relevant (using negative sentiment, using datasets from a given domain, etc.).

Recent work has started to move towards search with instructions, but this topic is still understudied with only a handful of papers~\citep{INSTRUCTOR,asai2022tart,muennighoff2024generative}. 
In particular, we find their use of instructions to be narrow:
instructions are typically short (fewer than 10 words) and repetitive (only one instruction per dataset \textit{e.g.}, \citet{INSTRUCTOR,asai2022tart,li2023angle,bge_embedding}).
Further, these works lack evaluation datasets that explicitly measure instruction following---instead focusing on standard \textit{ad-hoc} retrieval benchmarks. 

\begin{figure}
    \centering
    \includegraphics[width=\linewidth]{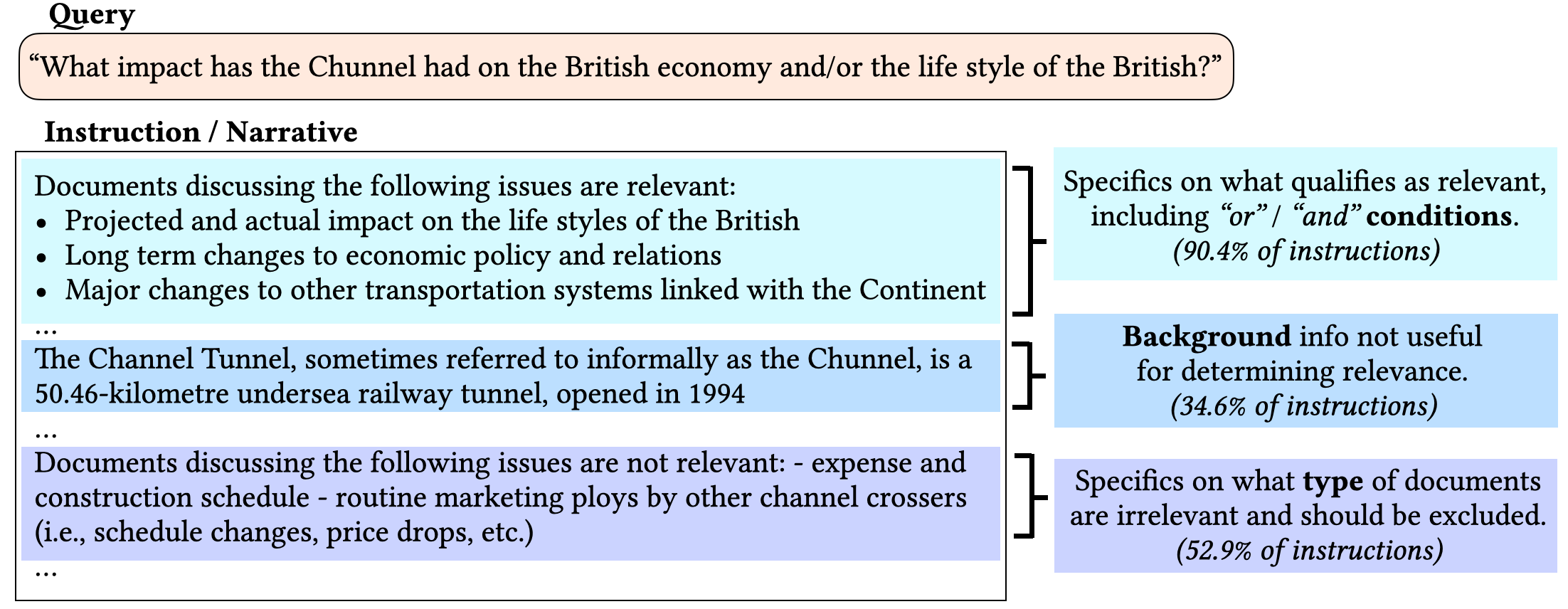}
    \vspace{-0.3em}
\caption{How do standard retrieval queries differ from instructions (or \textit{narratives})? Instructions contain more specific details about what is relevant, include less directly-relevant background information, and often have directives about what documents are \textit{not} relevant, using negation. \%'s are how often a certain type of content appears in the original TREC instructions used in \dataset.}
    \label{fig:narratives_diff}
\end{figure}

\begin{figure*}[t]
    \centering

    \includegraphics[width=0.9\linewidth,trim=0.2cm 0.2cm 0.4cm 1cm]{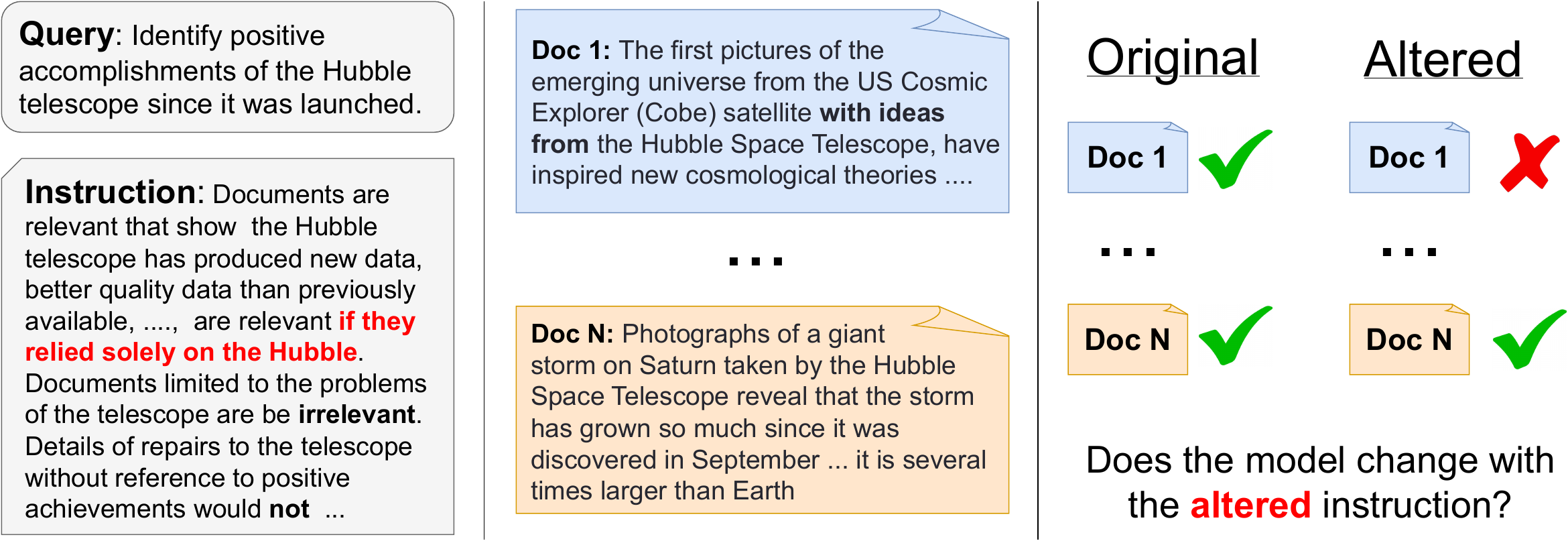}
    \caption{A visual depiction of the pairwise evaluation framework: models are evaluated on the query with the original instruction, and then on the query with   the \textcolor{red}{altered instruction}. If the model correctly understands the instructions, it will change which documents are relevant w.r.t. the alteration (right). Note that the real-world instructions (left) given to TREC annotators includes fine-grained details about what relevance is, as well as instructions containing negation (in \textbf{bold}).}
    \label{fig:teaser}
\end{figure*}

To address these gaps we introduce \dataset, which consists of (1) a benchmark that explicitly measures the instruction following ability of retrieval models, and (2) training data that includes diverse and realistic instructions. 
Our key intuition is to leverage instructions developed for \textit{professional annotators} of IR systems in order to study the capabilities of instruction-following IR models.
These instructions are used by annotators to judge document relevance for a given query.
Fortunately, the IR field is rich with such data, as these instructions---also known as \textit{narratives}---are created for all queries in any well-constructed IR dataset.
In particular, we use narratives developed for shared tasks at the Text REtrieval Conference\footnote{\href{http://trec.nist.gov}{\path{trec.nist.gov}}} (TREC).
These instructions are thorough and complex, including minute details about what makes a document relevant vs not-relevant.  
Thus if annotators can use these TREC instructions to annotate document relevance, so should instruction-following retrieval models (example query and instruction pairs are shown in Figures~\ref{fig:narratives_diff} and \ref{fig:teaser}).

We use three deeply-judged\footnote{\textit{i.e.}, that a large number of documents have been judged relevant or non-relevant, see Section~\ref{sec:trec} for more.} TREC collections as the basis of our evaluation set: TREC Robust 2004 \citep{voorhees2005trec}, TREC Common Core 2017 \citep{allan2017trec}, and TREC News 2021 \citep{soboroff2020trec}. 
These collections have been thoroughly annotated in order to evaluate recall in retrieval, with hundreds to thousands of documents judged as relevant or not-relevant.  
We take the instructions given to the professional annotators and alter them slightly, manually re-annotating the relevant documents. 
We then have paired instructions, which can be used to test how models respond to changing instructions; that is, we measure if models update the set of relevant documents to match to the new altered instructions.

As there are no existing methods to compare pairwise queries in IR, we develop a new evaluation framework to do so, measuring rank-wise score changes (which we call \metricrank) of documents given a pair of different instructions with the same query. 
Results on \dataset\ indicate that current models generally fail to follow instructions in retrieval unless they are 3B+ parameters or have not been trained for retrieval. Our analysis shows that these failures are due to two phenomena: (1) models are not used to long instructions, and (2) models use instruction for keyword search rather than as a definition of relevance.

To further progress in building retrieval models that can understand instructions, we build a training set of real-world human-used instructions and fine-tune a model on them (\dataset\--7B). Our results show marked improvement on \dataset\ for both standard IR metrics and for \metricrank, indicating a starting point for future progress on instruction following. 

In summary, we contribute the following: (1) a benchmark for evaluating instruction following in retrieval (\dataset) consisting of human annotations on top of three already highly-judged corpora, (2) analysis of why current models fail to understand instructions, and (3) training data for teaching retrieval models to follow instructions along with a new open-sourced IR model, \dataset\--7B, that can handle long instructions in IR.\footnote{Links to the code, data, and models are available at \url{https://github.com/orionw/FollowIR}}

\section{Related Work}
\subsection{TREC Conferences}
\label{sec:trec}
The United States National Institute of Science and Technology (NIST) created the TREC organization in 1993. Each year TREC sponsors many \textit{tracks}, or shared tasks, on a given dataset. These tracks range from a variety of topics: anywhere from standard ad-hoc retrieval on news \citep{soboroff2018trec,soboroff2021overview} to more complex domains such as legal retrieval \citep{baron2006trec,oard2008overview}, or retrieval-augmented generation/report-generation \citep{lawrie2023overview}.

As part of this process, NIST sponsors annotations for these collections. Typically, this is done by \textit{pooling} a set of results (\textit{runs}) from a variety of retrieval models and then annotating them in rank order until funding runs out. To help facilitate annotation, track organizers provide a \textit{narrative} (or instruction) for each query that will be given to the annotators---however, IR models are only ever given the query. As evaluating total recall would require annotating every document in the collection for every query (which is not feasible for collections with millions of documents), recall error is tested using post-hoc sampling and annotation. Although not every query and document pair can be evaluated, recall for queries is very high. We build off the rigorous evaluation done at TREC by using several of their collections to build \dataset.

\subsection{Instructions for LMs}
Instruction-following LMs have been popularized by models such as InstructGPT~\citep{instructGPT}, FLAN~\citep{wei2022finetuned}, and T0~\citep{sanh2022multitask}. 
They have become a large area of interest for the natural language processing community \citep{touvron2023llama,Jiang2023Mistral7,Groeneveld2024OLMoAT,Black2022GPTNeoX20BAO}. There has been much work in evaluating if they can generalize to new instructions \citep{weller2020learning,wang2022super,ouyang2022training}, if we can train them to follow instructions without human-annotated data \citep{wang2022self,qin2023toolllm}, and applying them to various domains \citep{zhao2021evaluation,singhal2023large,shaghaghian2020customizing}. As the IR community uses LMs in their pipelines, we seek to broaden the scope of IR to include instructions, aligning it with the broader NLP community.

\subsection{Instructions for Retrieval}
Using instructions in retrieval is a nascent area of exploration. \citet{INSTRUCTOR} and \citet{asai2022tart} were two of the earliest works that trained a retrieval model to use instructions along with the query. However, these instructions are typically very short, such as ``Retrieve a Wikipedia paragraph that answers this question."  Recent work incorporates instructions in smaller models \citep{bge_embedding,bge_m3,chen2024generalizing} as well as others which use Llama \citep{touvron2023llama,weller2023generative} or Mistral \citep{Jiang2023Mistral7} as the backbone of a larger retrieval model that can use instructions: GritLM \citep{muennighoff2024generative} trains Mistral to do both generation and embedding, while \citet{wang2023improving} uses Mistral for embeddings only.

Despite this flurry of activity, these efforts do not have an explicit instruction-related retrieval benchmark to evaluate on. Instead, they evaluate on standard retrieval benchmark suites such as MTEB \citep{muennighoff2022mteb} and BEIR \citep{thakur2021beir} which do not contain instructions. Thus, these newer instruction retrieval models hand-write a few instructions, where typically each instruction is applied to an entire dataset, irrespective of the query. This makes these instructions generic: focused only on the task format, format of the ``document" (paragraph, sentence, etc.), and the broad domain. Note that because of this, no current instructions contain any extra background information or negation \citep{Weller2024NevIRNI} which are commonly found in real-world instructions (see Figure~\ref{fig:narratives_diff} for an example of these differences).

In work concurrent to ours, \cite{oh2024instructir} also propose a dataset to evaluate instructions in retrieval models. Their dataset uses the MS MARCO collection~\citep{msmarco}, and differs in several crucial aspects: it only has one relevant document per query (\textit{e.g.}, sparsely judged), is GPT-4 generated and validated, focuses on the background of the user (``I am a school teaching looking for ..."), and evaluates using the lowest score over $N$ instructions for the same query (measuring robustness). In contrast, we use highly-judged corpora to ensure we can measure recall, use professionally generated instructions, have human-validated relevance judgements, propose a new paired evaluation protocol, and provide a training dataset and model for teaching instruction-following. 

\begin{table*}[t!]
\centering
\small
\begin{tabular}{l|cccc|ccc}
\toprule
Dataset & \# $Q$ & $|Q|$ & $|I|$ & Rel. D/Q & \# $Q$ & $|I|$ & Rel. D/Q \\
\midrule
TREC News '21 \citep{soboroff2020trec} & 50 & 15.3 & 40.1 & 50.1 & 32 & 46.9 & 19.2 \\
TREC Core '17 \citep{allan2017trec} & 50 & 16.6 & 44.0 & 180.0 & 20 &  53.5 & 32.7 \\
TREC Robust '04 \citep{voorhees2005trec} & 249 & 11.9 & 68.2 & 69.9 & 52 & 75.2 & 19.8 \\
\bottomrule
\end{tabular}
\caption{\dataset\ evaluation set statistics before (left) and after (right) annotation. We use a subset of the queries in three popular TREC tracks for variety in queries and documents. \textit{$|Q|$} is the word length of the queries and \textit{$|I|$} is the word length of the instructions. Rel. D/Q indicates the number of relevant annotated documents in the collection, excluding irrelevant annotations. As designed, there are less relevantly-judged documents in the \dataset\ portion (as the annotations change the relevance of documents on purpose for evaluation).
}
\label{tab:statistics}
\end{table*}

\section{Building \dataset}
We derive \dataset\ from three TREC collections: TREC News 2021 (derived from the Washington Post v4 corpus; \citealp{soboroff2020trec}), TREC Robust 2004 (from news articles in Disks 4 and 5 collections; \citealp{voorhees2005trec}), and TREC Common Core 2017 (from the New York Times Annotated corpus; \citealp{allan2017trec}). 
Each of these was professionally assessed to include hundreds of annotations per query (see Table~\ref{tab:statistics}), with 50-180 relevant documents per query on average (and many more not-relevant annotations).

Each of these TREC tracks includes instructions for the professional annotators that we now also give to the models. Although using these alone can provide some indication of how well models can follow instructions, it doesn't explicitly test their instruction following ability. To more carefully isolate this in our benchmark, we test whether models can respond to small changes in the instruction.

To accomplish this, we ask two expert annotators to modify the TREC instructions. However, doing this in a naive way would require re-annotating all the document judgements, a non-trivial task requiring immense annotation efforts.\footnote{NIST's  budget is \$1--2 million USD/year: \href{https://trec.nist.gov/pubs/2010.economic.impact.pdf}{\path{trec.nist.gov/pubs/2010.economic.impact.pdf}}} 
Instead, we task the annotators with making instructions \textit{more specific} by including additional constraints that narrow the relevance definition. 
These transformations cause some previously relevant documents to become non-relevant without introducing any new relevant documents from the pool.
Therefore, only those documents that were deemed relevant by the original TREC assessors need to be re-annotated.
This makes the annotation tractable, with only dozens or hundreds of documents to re-annotate per query instead of a collection of thousands.

We annotate a subset of the original TREC queries due to cost and overlap: we sample 50 queries from TREC Robust 2004 that do not overlap with TREC Common Core (as Common Core used 50 queries from Robust04 on a new collection), and 30 queries from TREC News 2021. Table~\ref{tab:statistics} shows dataset statistics of judged documents and the final benchmark size. 
Annotators were asked to change the instructions so that the number of relevant documents was cut roughly in half, thus including a sizeable number of changed relevance judgements.
We note that although the number of queries seems small by NLP standards, 30-50 queries is both effective \citep{webber2008statistical} and standard in the IR community due to the expense of careful annotation over many documents per query.

Due to differences in retriever quality, if we evaluate by searching over the full collection, each model will retrieve a different number of relevant documents. However, because we evaluate instruction following based on changing the document relevance, models that do poorly in the initial retrieval will have fewer documents which change relevance in the instruction-following evaluation. To rectify this, we instead turn to a reranking task where we include all relevant documents, and use a pool of five models\footnote{We use BM25, BGE-base, E5-base-v2, TART-Contriever, and INSTRUCTOR-xl.} to select the top non-relevant documents. To be able to freely distribute the data due to fair-use laws, we chunk the documents into 400-word passages with 200-word overlap and select the highest scoring passages using MaxP \citep{dai2019deeper}. This enables us to distribute our data, which we do by extending the MTEB evaluation framework \citep{muennighoff2022mteb}.

\subsection{Evaluation Metrics for \dataset}
Our benchmark provides two ways of measuring instruction following: (1) standard retrieval metrics when using the instructions with the queries and (2) pairwise evaluation of instruction following. For (1), we use typical IR evaluation metrics but use the instruction along with the query: these metrics are mean average precision (MAP) for Core17/Robust04 and normalized discounted cumulative gain at 5 (nDCG@5) for News21. For (2) we use our novel pairwise evaluation metric that measures the delta in scores when following the modified instructions instead of the original.\footnote{Note that we do not show standard retrieval results on the modified instruction's relevant document set, as standard retrieval scores cannot be directly compared across different query relevance annotations (\textit{qrels}).} 

Our new pairwise evaluation metric, \metricrank, measures rank-wise changes between queries. In developing this metric we had the following desiderata: it should compare the results of the original instruction to those of the new instruction, it should have a standardized range from worst possible change in instruction-following score (\textit{i.e.}, $-1$) to best possible instruction-following score (\textit{i.e.}, $1$) with an option for no change when using different instructions (\textit{i.e.}, $0$), and finally should take into account the document rank so that changes from rank 1 to rank 2 are more prominent than changes from rank 99 to 100. Given the above qualifications, we use the following equation applied to \textit{each} changed relevance document per query (where MRR is mean reciprocal rank, $R_{og}$ is the rank of the document when using the original instruction and $R_{new}$ is the new rank):
\begin{equation}
 \text{\metricrank} = 
  \begin{cases}
    \frac{MRR_{og}}{MRR_{new}} - 1 & \text{if $R_{og} > R_{new}$} \\[10pt]
   1 - \frac{MRR_{new}}{MRR_{og}} & \text{otherwise} 
  \end{cases}
\end{equation}
For the final score, we average first within a given query and then over all queries in the corpora---\textit{i.e.}, macro-averaging across queries, to handle the different number of relevant documents per query. 

\section{Evaluating Instruction Following}
In this section we describe the models we evaluate, their results on \dataset, and ablations performed to better understand the behavior of current models. 

\subsection{Evaluation Settings}
We evaluate a wide variety of IR models (trained with and without instructions), including neural models ranging from 100 million to 7 billion parameters. We evaluate on the original TREC instructions in the \dataset\ benchmark and then on the new instructions, showing both standard IR metrics and the new pairwise metric \metricrank. We group models into four categories:

\paragraph{No Instructions in Training} These  retrieval models did not see instructions in training and typically aren't given them: this includes Contriever \citep{izacard2021unsupervised}, E5 \citep{wang2022text}, MonoBERT \cite{nogueira2019multi}, MonoT5 \citep{nogueira2020document}, and BM25 \citep{robertson1995okapi}. 

\setlength\tabcolsep{5 pt} %
\begin{table*}[]
\small	
\centering
\resizebox{1.0\textwidth}{!}{
\begin{tabular}{ll|cc|cc|cc|cc}
\toprule
 & & \multicolumn{2}{c}{\textbf{Robust04}} & \multicolumn{2}{c}{\textbf{News21}} & \multicolumn{2}{c}{\textbf{Core17}} & \multicolumn{2}{c}{\textbf{Average}} \\ 
\multicolumn{2}{c}{\textbf{Model}} &  MAP & \metricrank & nDCG & \metricrank & MAP & \metricrank & Score & \metricrank \\
\midrule
\parbox[t]{3mm}{\multirow{7}{*}{\rotatebox[origin=c]{90}{
\parbox[c]{2cm}{\centering \scriptsize No-Instruction IR}}}}  
 & E5-base-v2 & 13.4 & \midpointgradientcell{-6.7}{-10.4}{13.7}{0}{neg}{pos}{\opacity}{0} & 20.9 & \midpointgradientcell{-2.0}{-2.8}{8.9}{0}{neg}{pos}{\opacity}{0} & 14.0 & \midpointgradientcell{-2.9}{-4.1}{16.5}{0}{neg}{pos}{\opacity}{0} & 16.1 & \midpointgradientcell{-3.9}{-4.4}{12.2}{0}{neg}{pos}{\opacity}{0} \\
  & Contriever & 19.7 & \midpointgradientcell{-6.1}{-10.4}{13.7}{0}{neg}{pos}{\opacity}{0} & 22.9 & \midpointgradientcell{-2.8}{-2.8}{8.9}{0}{neg}{pos}{\opacity}{0} & 15.3 & \midpointgradientcell{-2.5}{-4.1}{16.5}{0}{neg}{pos}{\opacity}{0} & 19.3 & \midpointgradientcell{-3.8}{-4.4}{12.2}{0}{neg}{pos}{\opacity}{0} \\
   & MonoBERT & 21.0 & \midpointgradientcell{-9.4}{-10.4}{13.7}{0}{neg}{pos}{\opacity}{0} & 25.1 & \midpointgradientcell{-0.8}{-2.8}{8.9}{0}{neg}{pos}{\opacity}{0} & 18.4 & \midpointgradientcell{-0.2}{-4.1}{16.5}{0}{neg}{pos}{\opacity}{0} & 21.5 & \midpointgradientcell{-3.5}{-4.4}{12.2}{0}{neg}{pos}{\opacity}{0} \\

 & BM25 & 12.1 & \midpointgradientcell{-3.1}{-10.4}{13.7}{0}{neg}{pos}{\opacity}{0} & 19.3 & \midpointgradientcell{-2.1}{-2.8}{8.9}{0}{neg}{pos}{\opacity}{0} & 8.1 & \midpointgradientcell{-1.1}{-4.1}{16.5}{0}{neg}{pos}{\opacity}{0} & 13.2 & \midpointgradientcell{-2.1}{-4.4}{12.2}{0}{neg}{pos}{\opacity}{0} \\

 & MonoT5-base & 15.7 & \midpointgradientcell{-6.2}{-10.4}{13.7}{0}{neg}{pos}{\opacity}{0} & 11.0 & \midpointgradientcell{+5.0}{-2.8}{8.9}{0}{neg}{pos}{\opacity}{0} & 12.2 & \midpointgradientcell{-4.1}{-4.1}{16.5}{0}{neg}{pos}{\opacity}{0} & 13.0 & \midpointgradientcell{-1.8}{-4.4}{12.2}{0}{neg}{pos}{\opacity}{0} \\
 & E5-large-v2 & 17.4 & \midpointgradientcell{-4.2}{-10.4}{13.7}{0}{neg}{pos}{\opacity}{0} & 24.3 & \midpointgradientcell{+0.9}{-2.8}{8.9}{0}{neg}{pos}{\opacity}{0} & 17.0 & \midpointgradientcell{+0.1}{-4.1}{16.5}{0}{neg}{pos}{\opacity}{0} & 19.6 & \midpointgradientcell{-1.1}{-4.4}{12.2}{0}{neg}{pos}{\opacity}{0} \\
 & MonoT5-3B & 27.3 & \midpointgradientcell{+4.0}{-10.4}{13.7}{0}{neg}{pos}{\opacity}{0} & 16.5 & \midpointgradientcell{+1.8}{-2.8}{8.9}{0}{neg}{pos}{\opacity}{0} & 18.2 & \midpointgradientcell{+1.8}{-4.1}{16.5}{0}{neg}{pos}{\opacity}{0} & 20.7 & \midpointgradientcell{+2.5}{-4.4}{12.2}{0}{neg}{pos}{\opacity}{0} \\

\midrule
 \parbox[t]{3mm}{\multirow{7}{*}{\rotatebox[origin=c]{90}{
\parbox[c]{2cm}{\centering \scriptsize Instruction-IR }}}}
 & TART-Contriever & 14.3 & \midpointgradientcell{-9.0}{-10.4}{13.7}{0}{neg}{pos}{\opacity}{0} & 21.8 & \midpointgradientcell{-3.0}{-3.0}{8.9}{0}{neg}{pos}{\opacity}{0} & 13.3 & \midpointgradientcell{-3.0}{-4.1}{16.5}{0}{neg}{pos}{\opacity}{0} & 16.5 & \midpointgradientcell{-5.0}{-5.0}{12.2}{0}{neg}{pos}{\opacity}{0} \\
  & INSTRUCTOR-base & 17.2 & \midpointgradientcell{-10.4}{-10.4}{13.7}{0}{neg}{pos}{\opacity}{0} & 22.1 & \midpointgradientcell{-1.8}{-2.8}{8.9}{0}{neg}{pos}{\opacity}{0} & 15.5 & \midpointgradientcell{-1.1}{-4.1}{16.5}{0}{neg}{pos}{\opacity}{0} & 18.3 & \midpointgradientcell{-4.4}{-4.4}{12.2}{0}{neg}{pos}{\opacity}{0} \\
& E5-mistral & 23.1 & \midpointgradientcell{-9.6}{-10.4}{13.7}{0}{neg}{pos}{\opacity}{0} & 27.8 & \midpointgradientcell{-0.9}{-2.8}{8.9}{0}{neg}{pos}{\opacity}{0} & 18.3 & \midpointgradientcell{+0.1}{-4.1}{16.5}{0}{neg}{pos}{\opacity}{0} & 23.1 & \midpointgradientcell{-3.5}{-4.4}{12.2}{0}{neg}{pos}{\opacity}{0} \\
 & BGE-base & 16.8 & \midpointgradientcell{-6.5}{-10.4}{13.7}{0}{neg}{pos}{\opacity}{0} & 20.0 & \midpointgradientcell{-0.1}{-2.8}{8.9}{0}{neg}{pos}{\opacity}{0} & 14.6 & \midpointgradientcell{-2.7}{-4.1}{16.5}{0}{neg}{pos}{\opacity}{0} & 17.1 & \midpointgradientcell{-3.1}{-4.4}{12.2}{0}{neg}{pos}{\opacity}{0} \\
  & INSTRUCTOR-xl & 19.7 & \midpointgradientcell{-8.1}{-10.4}{13.7}{0}{neg}{pos}{\opacity}{0} & 26.1 & \midpointgradientcell{-0.9}{-2.8}{8.9}{0}{neg}{pos}{\opacity}{0} & 16.8 & \midpointgradientcell{+0.7}{-4.1}{16.5}{0}{neg}{pos}{\opacity}{0} & 20.9 & \midpointgradientcell{-2.8}{-4.4}{12.2}{0}{neg}{pos}{\opacity}{0} \\
 & BGE-large & 17.5 & \midpointgradientcell{-7.8}{-10.4}{13.7}{0}{neg}{pos}{\opacity}{0} & 22.3 & \midpointgradientcell{+0.6}{-2.8}{8.9}{0}{neg}{pos}{\opacity}{0} & 15.0 & \midpointgradientcell{+0.1}{-4.1}{16.5}{0}{neg}{pos}{\opacity}{0} & 18.3 & \midpointgradientcell{-2.4}{-4.4}{12.2}{0}{neg}{pos}{\opacity}{0} \\
 & GritLM-7B & 28.6 & \midpointgradientcell{-1.7}{-10.4}{13.7}{0}{neg}{pos}{\opacity}{0} & 24.4 & \midpointgradientcell{-1.0}{-2.8}{8.9}{0}{neg}{pos}{\opacity}{0} & 20.8 & \midpointgradientcell{+2.6}{-4.1}{16.5}{0}{neg}{pos}{\opacity}{0} & 24.6 & \midpointgradientcell{-0.0}{-4.4}{12.2}{0}{neg}{pos}{\opacity}{0} \\
 & TART-FLAN-T5-xl & 24.6 & \midpointgradientcell{-0.7}{-10.4}{13.7}{0}{neg}{pos}{\opacity}{0} & 12.8 & \midpointgradientcell{+2.0}{-2.8}{8.9}{0}{neg}{pos}{\opacity}{0} & 17.0 & \midpointgradientcell{+2.8}{-4.1}{16.5}{0}{neg}{pos}{\opacity}{0} & 18.1 & \midpointgradientcell{+1.4}{-4.4}{12.2}{0}{neg}{pos}{\opacity}{0} \\

 \midrule
   \parbox[t]{3mm}{\multirow{3}{*}{\rotatebox[origin=c]{90}{
\parbox[c]{0.5cm}{\centering \scriptsize APIs}}}} 

 & OpenAI v3 Large & 27.2 & \midpointgradientcell{-5.8}{-10.4}{13.7}{0}{neg}{pos}{\opacity}{0} & 27.2 & \midpointgradientcell{-2.0}{-2.8}{8.9}{0}{neg}{pos}{\opacity}{0} & 21.6 & \midpointgradientcell{-0.2}{-4.1}{16.5}{0}{neg}{pos}{\opacity}{0} & 25.3 & \midpointgradientcell{-2.7}{-4.4}{12.2}{0}{neg}{pos}{\opacity}{0} \\
 & Cohere v3 English & 22.3 & \midpointgradientcell{-3.6}{-10.4}{13.7}{0}{neg}{pos}{\opacity}{0} & 28.3 & \midpointgradientcell{+0.2}{-2.8}{8.9}{0}{neg}{pos}{\opacity}{0} & 20.6 & \midpointgradientcell{+2.8}{-4.1}{16.5}{0}{neg}{pos}{\opacity}{0} & 23.7 & \midpointgradientcell{-0.2}{-4.4}{12.2}{0}{neg}{pos}{\opacity}{0} \\
 & Google Gecko & 23.3 & \midpointgradientcell{-2.4}{-10.4}{13.7}{0}{neg}{pos}{\opacity}{0} & 29.5 & \midpointgradientcell{+3.9}{-2.8}{8.9}{0}{neg}{pos}{\opacity}{0} & 23.2 & \midpointgradientcell{+5.4}{-4.1}{16.5}{0}{neg}{pos}{\opacity}{0} & 25.3 & \midpointgradientcell{+2.3}{-4.4}{12.2}{0}{neg}{pos}{\opacity}{0} \\

 \midrule
   \parbox[t]{3mm}{\multirow{5}{*}{\rotatebox[origin=c]{90}{
\parbox[c]{1.75cm}{\centering \scriptsize Instruct LMs}}}}
 & FLAN-T5-base & 6.4 & \midpointgradientcell{+5.3}{-10.4}{13.7}{0}{neg}{pos}{\opacity}{0} & 6.1 & \midpointgradientcell{-0.1}{-2.8}{8.9}{0}{neg}{pos}{\opacity}{0} & 6.5 & \midpointgradientcell{-3.3}{-4.1}{16.5}{0}{neg}{pos}{\opacity}{0} & 6.3 & \midpointgradientcell{+0.6}{-4.4}{12.2}{0}{neg}{pos}{\opacity}{0} \\
 & Llama-2-7B-chat & 6.3 & \midpointgradientcell{+2.0}{-10.4}{13.7}{0}{neg}{pos}{\opacity}{0} & 1.7 & \midpointgradientcell{+0.2}{-2.8}{8.9}{0}{neg}{pos}{\opacity}{0} & 5.4 & \midpointgradientcell{+2.8}{-4.1}{16.5}{0}{neg}{pos}{\opacity}{0} & 4.5 & \midpointgradientcell{+1.7}{-4.4}{12.2}{0}{neg}{pos}{\opacity}{0} \\
 & FLAN-T5-large & 14.7 & \midpointgradientcell{+3.9}{-10.4}{13.7}{0}{neg}{pos}{\opacity}{0} & 8.0 & \midpointgradientcell{+8.9}{-2.8}{8.9}{0}{neg}{pos}{\opacity}{0} & 11.4 & \midpointgradientcell{+1.3}{-4.1}{16.5}{0}{neg}{pos}{\opacity}{0} & 11.4 & \midpointgradientcell{+4.7}{-4.4}{12.2}{0}{neg}{pos}{\opacity}{0} \\
 & GritLM-Reranker & 9.7 & \midpointgradientcell{+6.1}{-10.4}{13.7}{0}{neg}{pos}{\opacity}{0} & 10.2 & \midpointgradientcell{+3.4}{-2.8}{8.9}{0}{neg}{pos}{\opacity}{0} & 9.8 & \midpointgradientcell{+8.6}{-4.1}{16.5}{0}{neg}{pos}{\opacity}{0} & 9.9 & \midpointgradientcell{+6.0}{-4.4}{12.2}{0}{neg}{pos}{\opacity}{0} \\
 & Mistral-7B-instruct & 23.2 & \midpointgradientcell{+12.6}{-10.4}{13.7}{0}{neg}{pos}{\opacity}{0} & 27.2 & \midpointgradientcell{+4.8}{-2.8}{8.9}{0}{neg}{pos}{\opacity}{0} & 19.7 & \midpointgradientcell{+13.0}{-4.1}{16.5}{0}{neg}{pos}{\opacity}{0} & 23.4 & \midpointgradientcell{+10.1}{-4.4}{12.2}{0}{neg}{pos}{\opacity}{0} \\
 & FollowIR-7B & 24.8 & \midpointgradientcell{+13.7}{-10.4}{13.7}{0}{neg}{pos}{\opacity}{0} & 29.6 & \midpointgradientcell{+6.3}{-2.8}{8.9}{0}{neg}{pos}{\opacity}{0} & 20.0 & \midpointgradientcell{+16.5}{-4.1}{16.5}{0}{neg}{pos}{\opacity}{0} & 24.8 & \midpointgradientcell{+12.2}{-4.4}{12.2}{0}{neg}{pos}{\opacity}{0} \\

\bottomrule
\end{tabular}
}
\caption{Evaluating instruction-following on \dataset. Introduced in this work, \metricrank\ is a pairwise evaluation metric measuring instruction following when instructions change, ranging from $-100$ to $100$ (higher is better).
Generally only models with over 3B parameters or instruction-tuned LMs that haven't been trained on retrieval tasks show success at following retrieval instruction.}
\label{tab:main}
\end{table*}

\paragraph{Instructions in IR Training} Most retrieval models using instructions received roughly one instruction per retrieval dataset, which generally defined the domain (\textit{e.g.}, ``Financial"), document size (sentence, passage, etc.), and task format. This includes INSTRUCTOR models \citep{INSTRUCTOR}, the bi-encoder TART model trained from Contriever \citep{asai2022tart}, the reranker TART trained from FLAN-T5 \citep{chung2022scaling}, E5 Mistral-Instruct \citep{wang2023improving}, and GritLM \citep{muennighoff2024generative}.  We also include BGE models \citep{bge_embedding} in this category, although they are trained with only one instruction total for each broad task (retrieval, clustering, etc.). 

\paragraph{API Models} We use three of the best performing API embedding models: Cohere's v3 English, Google's Gecko \citep{lee2024gecko} and OpenAI's Text-Embedding-v3-Large. It is mostly unknown what these models' training procedures were---including if they were trained on instructions or not---thus we place them in a distinct category. However, we note that Google's model did explicitly train with instructions, as mentioned in their technical report. 

\paragraph{Instruction-Tuned LMs} We also evaluate several instruction-tuned LMs to be used as rerankers, including FLAN-T5 \citep{chung2022scaling}, Llama v2 \citep{Touvron2023Llama2O}, and Mistral-Instruct-v0.2 \citep{Jiang2023Mistral7}. We evaluate these models in the same fashion as MonoT5 rerankers, comparing the true and false tokens. Note that these models were not trained on any retrieval-specific data.

\subsection{Results}
Table~\ref{tab:main} shows the main results on \dataset, with the standard IR score shown (either MAP or nDCG@5) as well as the pairwise evaluation metric, \metricrank.

\paragraph{No-Instruction IR Models} We see that the no-instruction models range widely in standard IR metrics (in terms of nDCG@5 and MAP) but generally have negative scores for \metricrank\ (up to $-3.9$). The only non-instruction model to score positively on average is MonoT5-3B (+2.5 \metricrank).

\paragraph{Instruction IR Models}  We again see that these models have generally negative scores, with the exception being GritLM (with scores averaging roughly zero) and TART-FLAN-T5-xl which has slightly positive scores for two of the three datasets (with an average of +1.4 \metricrank).

\begin{figure*}[t]
    \centering

    \includegraphics[width=0.9\linewidth,trim=0.0cm 0cm 0.0cm 2cm]{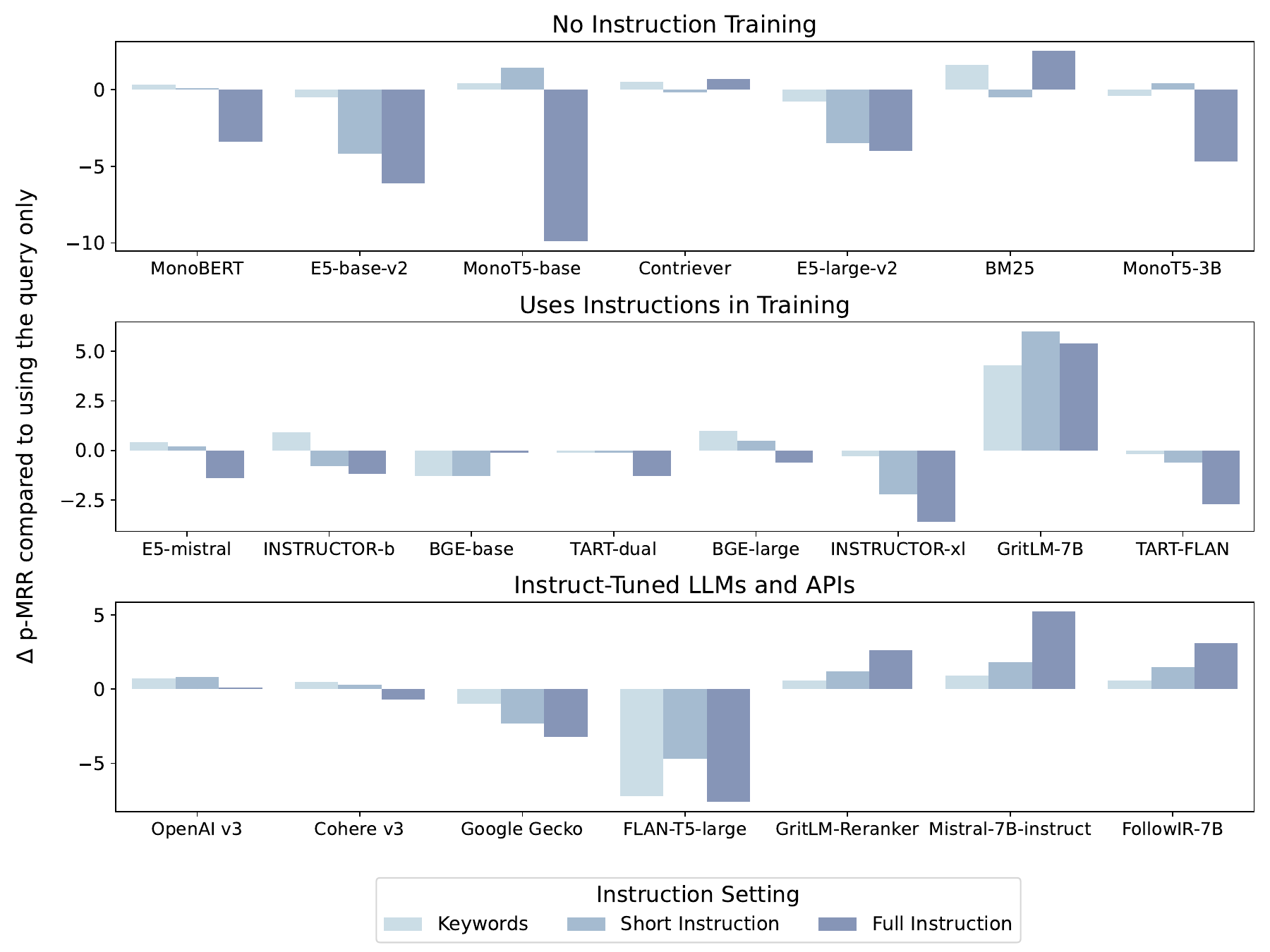}
    \caption{Score difference between using no instructions to using instructions formatted as keywords, short text, or the full text. While models that can correctly use instructions see gains with the additional information, most other models see decreasing performance as instruction length increases. 
    }
    \label{fig:ablation_ours}
\end{figure*}

\paragraph{API Models}
We see that the API models perform strongly in terms of standard IR metrics, with OpenAI's and Google's models performing the highest overall. However, Cohere's and OpenAI's models perform poorly at instruction-following with negative scores ($-0.2$ and $-2.7$ on average, respectively) whereas Google Gecko has positive scores (+2.3) likely a result of training on more instruction-focused data. 

\paragraph{Instruct-Tuned LMs} In contrast to the previous results, all instruction-tuned LMs show positive results for instruction following, although they have the widest range of performance using standard IR metrics (ranging from very poor scores to some of the higher scores). We see that the best performing model in this category is \dataset--7B, which we describe in more detail in Section~\ref{sec:model}. 

\paragraph{Overall} We see that the only models that show positive results at following instructions are either IR models with over 3B parameters or those that have been explicitly trained to follow instructions (\textit{e.g.} FLAN-T5), 
without any retrieval-specific supervision.
This aligns with work in the natural language processing community which has shown that the instruction-following ability improves with scale \citep{brown2020language} and supervised instruction-tuning \citep{longpre2023flan}.

\subsection{Analysis}
Why do so many models fail to correctly follow instructions when they do well on typical IR metrics such as nDCG and MAP? We answer this question by ablating several components that may impact results: (1) whether IR models are not used to text that cannot be used for simple keyword search (\textit{i.e.} instructions) and (2) whether they are unused to the length of the longer instructions (as current instruction IR models have been trained on much shorter instructions).

\setlength\tabcolsep{5 pt} %
\begin{table*}[]
\small	
\centering
\begin{tabular}{ll|rr|rr|rr}
\toprule
 & & \multicolumn{2}{c}{\textbf{SciFact}}  & \multicolumn{2}{c}{\textbf{NFCorpus}}  & \multicolumn{2}{c}{\textbf{FiQA}} \\ 
\multicolumn{2}{c}{\textbf{Model}} & OG & $\Delta$ w/Key. & OG & $\Delta$ w/Key. & OG & $\Delta$ w/Key. \\
\midrule
\parbox[t]{3mm}{\multirow{4}{*}{\rotatebox[origin=c]{90}{
\parbox[c]{1.3cm}{\centering \scriptsize No-Instruction}}}} & BM25 & 67.9 & -1.7 & 32.2 & -5.1 & 23.6 & -1.6  \\
& E5-base-v2 & 71.9 & -2.7 & 35.4 & -2.5 & 39.9 & -0.4  \\
& Contriever & 64.9 & +0.4 & 31.7 & +0.0 & 24.5 & -3.2  \\
& MonoT5-base & 73.1 & -0.6  & 35.6 & -0.9 & 41.2 & -0.3  \\
\midrule
\parbox[t]{3mm}{\multirow{6}{*}{\rotatebox[origin=c]{90}{
\parbox[c]{1.75cm}{\centering \scriptsize Uses Instruction}}}} 
& TART-Contriever & 67.6 & -0.3  & 33.4 & -5.3 & 31.8 & -0.4 \\
& INSTRUCTOR-base & 57.8 & +1.0 & 31.6 & -0.4 & 39.2 & -0.1  \\
& BGE-base & 73.2 & -0.5 & 35.5 & +0.0 & 40.8 & -2.3 \\
& TART-FLAN-xl & 74.2 & +1.6 & 33.9 & +0.4 & 39.6 & -0.3 \\
& INSTRUCTOR-xl & 62.4 & +0.2 & 36.0 & -0.6 & 46.9 & +0.8  \\
& E5-Mistral & 77.1 & -5.1  & 38.8  & +0.3 & 56.7 & -6.5  \\
\bottomrule
\end{tabular}

\caption{Ablation on BEIR benchmarks for models that do poorly with longer instructions, comparing their original short instructions vs domain keywords extracted from those instructions (see Appendix~\ref{app:extracted_beir} for a list). If models had learned to use the instructions correctly we would see a divergence between the behavior of instruct and non-instruct models, however, for both we see that using keywords  instead of the instruction results in comparable performance ($\pm$ one point). 
}
\label{tab:ablation_keywords}
\end{table*}

To test these, we compare the original query-only result to those where we additionally give the model either the full instruction, a shorter instruction, or keywords from the instruction. We gather these short instructions and keywords by prompting GPT-4-Turbo-1106 to generate them from the original full instruction (for TREC data) or otherwise use the original short instructions given by the authors of the model (for BEIR data).  For the full prompt text, please see Appendix~\ref{app:prompts}.

We show results for these ablations in Table~\ref{fig:ablation_ours}, where positive scores indicate that adding information improves the model while negative scores indicate a drop in performance. We see a consistent trend where models that did poorly on longer instructions perform better on keywords and shorter instructions than with the full instruction. However, models that are able to follow instructions see better results with the additional information, on average.

These results show that models are (1) using the instruction text as keywords (as performance is higher when using only keywords) and (2) are not able to utilize the extra information in the instructions (as they generally decrease in performance with this additional information).

We also confirm that these results hold on datasets outside of TREC collections and show results on three BEIR datasets: SciFact, NFCorpus, and FiQA.  We show in Table~\ref{tab:ablation_keywords} the original score (using the short instructions from their papers) and the change in score when using just keywords from the instruction (again extracted from GPT-4). We show results only for models which performed poorly for instruction-following.
We see that the scores for keywords vs the short instruction are generally similar, with most models seeing a change of around $\pm$ 1 point, except for the strongest of the non-instruction-following models, E5-Mistral, seeing a larger drop on some datasets.

\paragraph{Overall} We find overall (on both TREC and BEIR datasets) that models use instructions for keyword matching and are unused to longer instructions that may contain slightly less relevant words.

\section{Teaching Instruction Following}
\label{sec:model}
Is it possible to improve model performance in following instructions? We show that fine-tuning on a training set of longer instructions can provide a method for doing so. 

We start by gathering a training set to teach models. We collect all TREC narratives (\textit{i.e.}, instructions) from tasks not in \dataset, consisting of 1836 pairs of queries and narratives. However, we note that this does not provide any positive or negative documents for fine-tuning.

In order to obtain documents for training, we prompt GPT-3.5-Turbo-1106 to generate relevant and not-relevant documents, generating roughly two relevant and non-relevant instances per query. However, these synthetic documents are noisy and contains errors w.r.t. the labels---to remedy this, we perform a round of filtering and use the best performing open-source model from Table~\ref{tab:main} (Mistral-7B-Instruct-v0.2) to score each of the generated documents according to the instruction.  We then filter the documents according to whether Mistral correctly predicts the generated label, and finally balance the relevant and non-relevant samples, choosing only one relevant and non-relevant document per query. Our total is $\sim$1800 training instances on $\sim$1200 unique query/instructions pairs. 

We then train our instruction-following model, \dataset\--7B, by fine-tuning Mistral-7B-Instruct-v0.2 on our data using the Llama-Factory framework \citep{llama-factory} with LoRA \citep{hu2021lora}. Full training hyperparameter details are found in Appendix~\ref{app:hyperparameters}.

\begin{wrapfigure}[13]{r}{0.5\textwidth}
    \vspace{-10pt}
    \centering
    \setlength\tabcolsep{5 pt} %
    \begin{tabular}{lr}
    \toprule
    Model & Robustness@10 \\
    \midrule
    BM25 & 26.9 \\
    TART-Contriever & 47.5 \\
    RepLLaMa & 52.6 \\
    E5-Mistral & 55.4 \\
    \midrule
    Mistral-7B-instruct & 35.3 \\
    FollowIR-7B & 71.5 \\
    \bottomrule
    \end{tabular}
    \caption{
        Performance on the InstructIR benchmark using their ``Robustness@10" scores, e.g. the min nDCG@10 score across 10 instructions. Upper portion is bi-encoders while lower is rerankers.
        \label{tab:instructir}
    }
    \label{fig:example}
    \vspace{-1em}
\end{wrapfigure}

When we evaluate this model on \dataset\ (Table~\ref{tab:main}), we find that the scores consistently improve. Compared to the original Mistral-7B-Instruct-v0.2, our model improves on both standard IR metrics (+6.0\% relative improvement) and on instruction following (+20.8\% relative improvement). We also show that this improvement holds on the concurrent InstructIR dataset (Table~\ref{tab:instructir}), where FollowIR-7B scores double the base Mistral-7B scores (71.5 Robustness@10 vs 35.3) and is the top scoring model overall. Thus, we can see that it is possible to train IR models to be better instruction followers.

\section{Conclusion}
Despite the use of LMs as the backbone of neural retrieval models, most existing IR models do not take instructions that define document relevance. 
Further, there is no existing resource that measures how well retrieval models can follow instructions. 
We build a new benchmark that explicitly measures the instruction following ability of retrieval models and find that nearly all retrieval models do not follow instructions, with the exception of larger models (3B+ parameters) or instruction-tuned LMs that typically are not used for retrieval. 
However, we show that it is possible to improve their instruction following ability, and build and release a training corpus for teaching retrieval models to follow instructions.  
Our new model, \dataset\--7B, shows improvement on both standard retrieval metrics as well as in instruction following.
We hope that these resources will allow the community to develop more capable instruction-following retrieval models that can quickly adapt to a relevance definition given flexible natural language text.

\section{Limitations}
\paragraph{Reranking vs Full Retrieval} As our setup for evaluating instruction following requires evaluating the documents which changed relevance, we cannot use the full collection for retrieval (as each retriever finds different relevant documents by design). Further, due to licensing restrictions, we cannot distribute the full corpora from the TREC tracks---thus we distribute passages due to fair use laws. However, we show full corpus retrieval results for a subset of models in Appendix~\ref{app:full_retrieval} and note similar trends in terms of the lack of instruction following. 

\paragraph{Possible Errors} Our work is built on the TREC document collections and judgements, as well as new annotation efforts. We do not check for potential errors in the TREC annotations, and our newly gathered annotations may have small errors. Despite these caveats, we see that our dataset still provides a useful evaluation setup for measuring instruction following.

\bibliography{colm2024_conference}
\bibliographystyle{colm2024_conference}

\appendix
\section{Hyperparameters for Fine-Tuning Mistral}
\label{app:hyperparameters}

\section{Hyperparameters for Inference}
\label{app:hyperparameters_inferece}
We use default parameters for inference, taken from the original code of the authors of the papers we use (from their MTEB evaluations).

\section{Full Retrieval Results}
\label{app:full_retrieval}
In Table~\ref{tab:full_collection} we show results for models searching on the full collections of the TREC tasks included in \dataset. Note that because each model retrieves different relevant documents, the instruction-following evaluation has a different set of instances that each model is evaluated on (as it can only be evaluated on documents it retrieved that then become not-relevant).

\setlength\tabcolsep{5 pt} %
\begin{table*}[t!]
\small	
\centering
\begin{tabular}{ll|rr|rr|rr}
\toprule
 &  & \multicolumn{2}{c}{\textbf{Robust04}} & \multicolumn{2}{c}{\textbf{News21}} & \multicolumn{2}{c}{\textbf{Core17}} \\ 
 & & \multicolumn{2}{c}{(mAP)} & \multicolumn{2}{c}{(nDCG@5)} & \multicolumn{2}{c}{(mAP)} \\
\multicolumn{2}{c}{\textbf{Model}} &  OG & $\Delta$ & OG & $\Delta$ & OG & $\Delta$ \\
\midrule
\parbox[t]{3mm}{\multirow{3}{*}{\rotatebox[origin=c]{90}{
\parbox[c]{1cm}{\centering \scriptsize No \\ Instruct}}}}
& BM25 & 21.4 & -1.2 & 30.1 & +5.3 & 16.8 &  -0.2 \\
& E5-base-v2 & 22.7 & -7.0 & 33.6 & +1.8 & 19.7  & -3.0 \\
& Contriever & 19.2 & -7.7 & 22.5 & +9.0 & 22.6 & -7.6 \\
\midrule
\parbox[t]{3mm}{\multirow{4}{*}{\rotatebox[origin=c]{90}{
\parbox[c]{1.5cm}{\centering \scriptsize Uses \\ Instruct}}}} 
& TART-Contriever & 25.5 & -10.1 & 40.0 & -5.0 & 22.6 & -7.6 \\
& BGE-base & 23.6 & -3.1 & 36.5 & -7.8 & 23.0 & -2.1 \\
& INSTRUCTOR-base & 22.5  & -2.2 & 33.3 & -2.8 & 20.0 & -0.2 \\
& INSTRUCTOR-XL & 30.4 & -3.1 & 38.1 & -0.1 & 29.9 & -2.8 \\

\bottomrule
\end{tabular}
\caption{\dataset\ scores on the full retrieval collection (thus rerankers are not included). As the base score is different, there are different numbers of relevant documents they are being evaluated on for \metricrank. Thus, we only report the original (no-instruction) score and the delta when using the TREC instructions. We note that it shows similar results to the main text -- retrieval models are not effectively using instructions and see performance degradations with longer text.}
\label{tab:full_collection}
\end{table*}

\section{Keywords used for BEIR experiments}
\label{app:extracted_beir}
GPT-4-Turbo-1106 extracted the following keywords (Table~\ref{tab:keywords}) from the instructions these models used, which generated the results in Table~\ref{tab:ablation_keywords}. 

\setlength\tabcolsep{5 pt} %
\begin{table*}[t!]
\small	
\centering
\begin{tabular}{lrr}
\toprule
Model & Dataset & Keywords \\
\midrule
BM25/Contriever/E5/MonoT5 & FiQA & Finance web \\
BM25/Contriever/E5/MonoT5 & SciFact & science paper verify \\
BM25/Contriever/E5/MonoT5 & NFCorpus & medicine relevant \\
\midrule
TART-dual & FiQA & financial web \\
TART-dual  & SciFact & scientific paper verify \\
TART-dual  & NFCorpus & scientific paper paragraph \\
INSTRUCTOR-base & FiQA & financial supporting: \\
INSTRUCTOR-base & SciFact & scientific supporting passage: \\
INSTRUCTOR-base & NFCorpus & medicine relevant \\
BGE-base & FiQA & relevant passages: \\
BGE-base & SciFact & relevant passages: \\
BGE-base & NFCorpus & relevant passages: \\
INSTRUCTOR-xl & FiQA & finance supporting:\\
INSTRUCTOR-xl & SciFact & scientific supporting passages: \\
INSTRUCTOR-xl & NFCorpus & nutrition facts public medical:  \\
E5-Mistral & FiQA & financial replies \\
E5-Mistral  & SciFact & scientific \\
E5-Mistral  & NFCorpus & retrieve relevant \\
TART-T5-FLAN-xl & FiQA & financial web \\
TART-T5-FLAN-xl  & SciFact & scientific paper verify \\
TART-T5-FLAN-xl & NFCorpus & Scientific paper paragraph \\
\bottomrule
\end{tabular}
\caption{Keywords used for the BEIR keyword analysis. Note that non-instruction models received the keywords used in INSTRUCTOR-base and TART-dual (as shown in the table).\label{tab:keywords}}
\end{table*}

\section{Prompts Used}
\label{app:prompts}
We use these prompts for generating the short instructions, the keywords, and the synthetic documents. The examples used in the prompt for the ``Full Instructions to Short Instructions" prompt were partially created by the authors, as only the short instructions were provided by TART/INSTRUCTOR.

\begin{tcolorbox}[colback=white,colframe=black,title=Synthetic Document Creation]
I need you to annotate some data for my business and it is super important that you follow instructions precisely or you will be fired.\\

Given a Google search query and instructions regarding what makes a document relevant, I need you to write two documents: one that would be relevant and one that would not.\\

Search: TITLE\_HERE\\
Instructions: NARRATIVE\_HERE\\

I need some different options to choose from, so give me three **different** options for both a relevant document and an irrelevant document. They should be **long** paragraph-sized documents (\(\sim\)300 words each), one on each line. If there is no negation in the instructions, your irrelevant document should be slightly off topic:
\end{tcolorbox}

\begin{tcolorbox}[colback=white,colframe=black,title=Short Instructions to Keywords]
I have instructions that are specific to a style of retrieval, but I want you to instead just focus on the relevant keywords that are in these instructions. Your job is to return a list of these keywords that are relevant in the query. There are probably one or two relevant keywords to extract only.

\medskip

\noindent \#\# Examples\\
\#\#\# Example 1:\\
Instruction: Help me to find a highly related PubMed paper to answer this question.\\
Keywords: ["PubMed"]\newline

\smallskip

\noindent \#\#\# Example 2:\\
Instruction: I want to find an answer for this Trivia question. Can you find some paragraphs that provide evidence from Wikipedia?\\
Keywords: ["Trivia", "Wikipedia"]\newline

\smallskip

\noindent \#\#\# Example 3:\\
Instruction: Check if a Quora question is duplicated with this question.\\
Keywords: ["Quora", "duplicated"]\newline

\smallskip

\noindent \#\#\# Example 4:\\
Instruction: I want to find a related question asked in StackExchange. Can you find one for me?\\
Keywords: ["related", "StackExchange"]\newline

\medskip

\noindent \#\# Your turn\\
Instruction: FILL\_TEXT\_HERE\\
Keywords (either one or two keywords, that are not "documents", "questions", "answer", or "articles"): 
\end{tcolorbox}

\begin{tcolorbox}[colback=white,colframe=black,title=Full Instructions to Short Instructions]
I have instructions that are specific to a question, but I need your help abstracting them to a general task format that I can give to someone else.  I need you to turn them into an abstract command that just describe the general abstract task instead (e.g., where the data is from, what the type of document looks like). It is crucial that you read and follow these instructions, you will get a large bonus if you are successful (\$200).\newline

The abstract command should only mention the **task format**.   Do **not** refer to any entities or specific text in the original instruction. Your response should be around 10 words. The command should be as if you were speaking to another human.\newline

\#\# Examples\newline
\#\#\# Example 1:\newline
Original Instruction: A relevant document would provide information about the whole blood-base perfusate and whether or not it provides superior preservation of myocardial function during ex vivo heart perfusion. This may include research experiments, commentary, or survey/review papers. Information about whole blood-base perfusate alone is not relevant, unless it also mentions it's effect on myocardial function during ex vivo heart perfusion.\newline
Abstract Command: Help me to find a highly related PubMed paper to answer this question.\newline

\#\#\# Example 2:\newline
Original Instruction: A relevant document will contain information that about the right of way in international waters that can be used to determine who should have the right of way in a given situation. For example, it should contain instances about who is at fault in an accident, if it depends on the size of the boat, or details about how this differs according to nationality. Especially relevant are documents describing who is at fault in a crash situation.\newline
Abstract Command: Retrieve a Wikipedia paragraph that answers this question. \newline

\#\#\# Example 3:\newline
Original Instruction: A relevant instance will be a question that is semantically equivalent to the query given.  For example, it may contain different lexical words or be a paraphrase of the other, but the underlying meaning will be the same. If the instance is not semantically the same as the query, it is irrelevant.\newline
Abstract Command: Check if a Quora question is duplicated with this question.\newline

\#\#\# Example 4:\newline
Original Instruction: A relevant document would include details about the timing of medicare and what age patients can start using it for healthcare.  It may include information about laws, insurance, or other details that describe the age the medicare begins. Less relevant are documents talking about potential laws or facts about medicare that do not answer the question of what age medicare begins. Just the mention of an age and when to start medicare would be relevant.\newline
Abstract Command: I want to know the answer to the question. Can you find good evidence on the web?\newline

Now you can easily see that the abstract command is vague and describes only a short command about how to get the information you need. Follow this exactly---do not reference specifics (like in the above, "international waters" and "medicare" are not included in the abstract command). You should instead keep the abstract command vague and well, abstract about the task only. Use the word "question".\newline

\#\# Your turn\newline
Original Instruction: FILL\_TEXT\_HERE\newline
Abstract Command (remember to use the word "question"):
\end{tcolorbox}

\end{document}